\documentclass{amsart}

\usepackage{amssymb}

\usepackage[english,francais]{babel}

\newtheorem{e-proposition}[theorem]{Proposition}

\newtheorem{e-definition}[theorem]{Definition\rm}

\newtheorem{theoreme}{Th\'eor\`eme}[section]

\newtheorem{proposition}[theoreme]{Proposition}

\def\og{\leavevmode\raise.3ex\hbox{$\scriptscriptstyle\langle\!\langle$~}}
\def\fg{\leavevmode\raise.3ex\hbox{~$\!\scriptscriptstyle\,\rangle\!\rangle$}}

\def \Z {\mathbb Z}
\def \R {\mathbb R}
\def \C {\mathbb C}

\DeclareMathOperator{\SpN}{\mathrm{Sp}_{\mathrm{N}}(\R)}
\DeclareMathOperator{\spN}{\mathfrak{sp}_{\mathrm{N}}(\R)}

\DeclareMathOperator{\dd}{\mathrm{d}}
\DeclareMathOperator{\dlO}{d_{\log\, \mathcal{O}}}

\title[Anderson \`a valeurs matricielles]{Exposants de Lyapounov pour un mod\`ele d'Anderson
\`a valeurs matricielles}
\author{Hakim Boumaza}
\email{boumaza@math.jussieu.fr}
\address{Keio University, Department of Mathematics\\
Hiyoshi 3-14-1\\
Kohoku-ku 223-8522\\
Yokohama\\
Japan\\
}

\begin{document}
\maketitle
\begin{abstract}
% resume en francais, et apres l'abstract en anglais, qui
%    commence avec le titre en gras.
\selectlanguage{francais}
Nous pr\'esentons un r\'esultat d'absence de spectre absolument continu dans un intervalle de $\R$ pour un op\'erateur de Schr\"odinger al\'eatoire continu et \`a valeurs matricielles agissant sur $L^2(\R)\otimes \C^N$ pour $N\geq 1$ arbitraire. Pour cela nous prouvons l'existence d'un intervalle d'\'energies sur lequel a lieu la s\'eparabilit\'e et la stricte positivit\'e des $N$ exposants de Lyapounov positifs de l'op\'erateur. La m\'ethode suivie, bas\'ee sur le formalisme de F\"urstenberg et un r\'esultat de th\'eorie des groupes d\^u \`a Breuillard et Gelander, permet une construction explicite de l'intervalle d'\'energie recherch\'e.

\vskip 0.5\baselineskip
\end{abstract}

%-------------------------------------------------------------%
% texte principal
%-------------------------------------------------------------%
\selectlanguage{francais}
%-------------------------------------------------------------%
\section{Introduction}\label{secintro}

\noindent Pour les mod\`eles d'Anderson dans une bande continue du plan $\R \times [0,1]$, la question de la localisation \`a toutes les \'energies reste une question ouverte. Un tel mod\`ele est repr\'esent\'e par un op\'erateur al\'eatoire aux d\'eriv\'ees partielles de la forme $H=-\Delta + V_{\omega}$ agissant sur $L^2(\R \times [0,1])$ avec conditions de Dirichlet aux bords de la bande, $\R \times \{0 \}$ et $\R \times \{ 1\}$.  Le symbole $\Delta$ d\'esigne le laplacien continu en dimension $2$ et $V_{\omega}$ est une fonction sur $\R \times [0,1]$. Pour \'etudier l'op\'erateur $H$, l'id\'ee est d'op\'erer une discr\'etisation dans la direction o\`u la bande est de longueur finie. Cela permet de ramener le probl\`eme initial d'\'equation aux d\'eriv\'ees partielles \`a l'\'etude d'un syst\`eme diff\'erentiel ordinaire. Nous \'etudions donc un op\'erateur d'Anderson continu, unidimensionnel et \`a valeurs matricielles de la forme $H_N=-\frac{\mathrm{d}^2}{\mathrm{d}x^2}\otimes I_N + V_N(\omega)$ o\`u $I_N$ est la matrice identit\'e de taille $N\geq 1$ et $V_N(\omega)$ une fonction \`a valeurs dans les matrices sym\'etriques r\'eelles d\'ependant de param\`etres al\'eatoires. L'objectif est d'obtenir la localisation d'Anderson pour $H_N$ pour tout $N$ puis d'\'etudier s'il est possible d'obtenir la localisation d'Anderson pour $H$ en consid\'erant la limite lorsque $N$ tend vers l'infini.

\noindent Pour prouver la localisation d'Anderson sur un intervalle d'\'energies pour un op\'erateur de la forme de $H_N$, la premi\`ere \'etape est de prouver la s\'eparabilit\'e des exposants de Lyapounov associ\'es \`a $H_N$ sur cet intervalle, comme cela est fait dans \cite{KLS90} ou \cite{DSS02}. Dans \cite{boumaza-mpag} nous avions d\'ej\`a prouv\'e, dans le cas $N=2$, l'existence d'intervalles d'energies sur lesquels les exposants de Lyapounov associ\'es \`a un op\'erateur d'Anderson continu \`a valeurs matricielles \'etaient s\'epar\'es. Le but de cet article est de pr\'esenter un r\'esultat de s\'eparabilit\'e des exposants de Lyapounov associ\'es \`a un op\'erateur $H_N$ pour $N\geq 1$ arbitraire. Pour d\'emontrer un tel r\'esultat, nous aurons recours \`a un crit\`ere de densit\'e de sous-groupes de groupes de Lie semi-simples d\^u \`a Breuillard et Gelander (\cite{BG03}), suivant la m\^eme m\'ethode que dans \cite{boumaza-mpag}. En effet, la d\'emarche adopt\'ee ici est d'\'etudier la densit\'e du groupe de F\"urstenberg associ\'e \`a $H_N$ (\emph{i.e} le sous-groupe du groupe symplectique $\SpN$ engendr\'e par les matrices de transferts associ\'ees \`a $H_N$) dans $\SpN$. Cela permet d'obtenir aussi la r\'egularit\'e h\"old\'erienne des exposants de Lyapounov et de la densit\'e d'\'etats int\'egr\'ee de $H_N$ sur tout intervalle compact d'\'energie o\`u les exposants de Lyapounov sont s\'epar\'es (voir \cite{boumaza-rmp}). La r\'egularit\'e h\"old\'erienne de la densit\'e d'\'etats int\'egr\'ee de $H_N$ est une \'etape importante en vue d'appliquer un sch\'ema d'analyse multi-\'echelle pour prouver la localisation d'Anderson pour $H_N$ (voir \cite{stollmann, klein-msa}).

\section{Mod\`ele et r\'esultats}

\noindent Dans le pr\'esent article nous \'etudions le mod\`ele d'Anderson suivant :
\begin{equation}\label{modele}
H_{\ell}(\omega)=-\frac{\dd^{2}}{\dd x^{2}}\otimes I_{\mathrm{N}}+ V_{0} +\sum_{n\in \Z} \left(
\begin{matrix}
c_1 \omega_{1}^{(n)} \mathbf{1}_{[0,\ell]}(x-\ell n) & & 0\\ 
 & \ddots &  \\
0 & & c_N \omega_{N}^{(n)} \mathbf{1}_{[0,\ell]}(x-\ell n)\\ 
\end{matrix}\right)
\end{equation}
agissant sur $L^2(\R)\otimes \C^N$. On suppose que $N\geq 1$ est un entier, $I_{\mathrm{N}}$ est la matrice identit\'e d'ordre $N$, chaque $c_i$ est dans $\R^*$ et $\ell>0$. Pour $i\in \{ 1,\ldots,N\}$, les  $(\omega_{i}^{(n)})_{n\in \Z}$ sont des suites de variables al\'eatoires ind\'ependantes et identiquement distribu\'ees sur $(\Omega,\mathcal{A},\mathsf{P})$ de loi commune $\nu$ telle que $\{ 0,1\} \subset \mathrm{supp}\; \nu$. Enfin, $V_0$ est l'op\'erateur de multiplication par la matrice tridiagonale $V_0$ ayant une diagonale nulle et tous les coefficients de sa surdiagonale et de sa sous-diagonale \'egaux \`a $1$. Le param\`etre al\'eatoire $\omega$ est une variable al\'eatoire sur l'espace produit $(\otimes_{n\in \Z} \Omega^{\otimes N}, \otimes_{n\in \Z} \mathcal{A}^{\otimes N}, \otimes_{n\in \Z} \mathsf{P}^{\otimes N})$. On note pour tout $n\in \Z$, $\omega^{(n)}=(\omega_1^{(n)},\ldots,\omega_N^{(n)})$ qui est de loi $\nu^{\otimes N}$. Le param\`etre $\ell>0$ peut \^etre interpr\'et\'e comme une longueur d'interaction. On remarque que $H_{\ell}(\omega)$ est une perturbation born\'ee de l'op\'erateur $-\frac{\dd^{2}}{\dd x^{2}}\otimes I_{\mathrm{N}}$, il est donc autoadjoint sur l'espace de Sobolev $H^2(\R)\otimes \C^N$. 

\noindent Notre r\'esultat de s\'eparabilit\'e des exposants de Lyapounov de $H_{\ell}(\omega)$ est le suivant.

\begin{theoreme}\label{thm1}
Soit $N\geq 1$. Il existe $\ell_C=\ell_C(N) >0$ tel que pour tout $\ell<\ell_C$, il existe un intervalle compact $I=I(N,\ell)\subset \R$ (ne d\'ependant que de $\ell$ et de $N$ et dont la longueur tend l'infini lorsque $\ell$ tend vers $0$) tel que les $N$ exposants de Lyapounov positifs $\gamma_1(E),\ldots,\gamma_N(E)$ de $H_{\ell}(\omega)$ v\'erifient 
\begin{equation}\label{seplyapthm1}
\forall E\in I,\quad \gamma_{1}(E)>\cdots > \gamma_{N}(E)>0.
\end{equation}
En particulier, $H_{\ell}(\omega)$ n'a pas de spectre absolument continu dans $I$.  
\end{theoreme}

\section{Principe de la preuve du th\'eor\`eme \ref{thm1}}

\noindent Nous commen\c{c}ons par introduire les matrices de tranfert de l'op\'erateur $H_{\ell}(\omega)$. Soit $E\in \R$. La matrice de transfert de $\ell n$ \`a $\ell (n+1)$ de $H_{\ell}(\omega)$ est d\'efinie par la relation
\begin{equation}\label{mat_transfert_def}
\left( \begin{array}{c} 
u(\ell(n+1)) \\
u'(\ell(n+1))
\end{array} \right)= T_{\omega^{(n)}}(E) \left( \begin{array}{c}
u(\ell n) \\
u'(\ell n)
\end{array} \right)
\end{equation}
o\`u $u:\R \to \C^N$ est solution du syst\`eme diff\'erentiel de second ordre $H_{\ell}(\omega)u=Eu$. On introduit alors pour tout r\'eel $E$ le groupe de F\"urstenberg de $H_{\ell}(\omega)$ :
\begin{equation}\label{def_group_furst}
G(E)= \overline{<T_{\omega^{(0)}}(E)|\ \omega^{(0)} \in \mathrm{supp}\; \nu^{\otimes N}>} \supset \overline{<T_{\omega^{(0)}}(E)|\ \omega^{(0)} \in \{ 0,1 \}^N>}. 
\end{equation}

\noindent En vertu d'un th\'eor\`eme d\^u \`a Gol'dsheid et Margulis (voir \cite{GM89, stolzboumaza}), pour prouver que pour un r\'eel donn\'e $E$ les exposants de Lyapounov sont s\'epar\'es, il suffit de prouver que $G(E)$ est Zariski-dense dans $\SpN$. En fait nous allons prouver un r\'esultat plus fort.

\begin{proposition}\label{prop1}
Il existe $\ell_C$ et $I$ comme voulus au th\'eor\`eme \ref{thm1} tels que pour tout $E\in I$, $G(E)=\SpN$. 
\end{proposition}

\noindent Pour cela nous utilisons le r\'esultat suivant de th\'eorie des groupes d\^u \`a Breuillard et Gelander.

\begin{theoreme}[Breuillard et Gelander, \cite{BG03}]\label{thm_BG}
Si $G$ est un groupe de Lie connexe r\'eel semi-simple, d'alg\`ebre de Lie $\mathfrak{g}$, alors il existe un voisinage de l'identit\'e $\mathcal{O} \subset G$, sur lequel $\log=\exp^{-1}$ est un diff\'eomorphisme et tel que $g_{1},\ldots,g_{m}\in \mathcal{O}$ engendrent un sous-groupe dense dans $G$ lorsque $\log(g_{1}),\ldots, \log(g_{m})$ engendrent $\mathfrak{g}$. 
\end{theoreme}

\noindent Ce th\'eor\`eme nous donne le plan de la suite de la preuve. Tout d'abord nous allons calculer explicitement les matrices de tranfert $T_{\omega^{(0)}}(E)$ pour $\omega^{(0)}\in \{ 0,1\}^N$. Nous prouvons alors qu'il existe $\ell_C>0$ ne d\'ependant que de $N$ tel que pour tout $\ell <\ell_C$, $\ell>0$, il existe un intervalle compact $I(N,\ell)$ de $\R$ tel que pour tout $E\in I(N,\ell)$, $T_{\omega^{(0)}}(E) \in \mathcal{O}$ pour tout  $\omega^{(0)}\in \{ 0,1\}^N$. Ici, $\mathcal{O}$ est le voisinage de l'identit\'e donn\'e par le th\'eor\`eme \ref{thm_BG} pour $G=\SpN$. Ensuite, pour $\ell < \ell_C$, nous calculons les logarithmes des matrices $T_{\omega^{(0)}}(E)$ et nous prouvons qu'ils engendrent l'alg\`ebre de Lie $\spN$ de $\SpN$.
\vskip 3mm

\noindent Nous commen\c{c}ons par donner l'expression des matrices de transfert. Posons 
\begin{equation}\label{expr_mat_transfert}
M_{\omega^{(0)}}(E)=V_0 + \mathrm{diag}(c_1 \omega_1^{(0)}-E,\ldots, c_N \omega_N^{(0)}-E).
\end{equation}
\noindent Alors, si on note 
\begin{equation}\label{expr_mat_transfert_X}
X_{\omega^{(0)}}(E)=\left( \begin{array}{cc}
0 & I_{\mathrm{N}} \\
M_{\omega^{(0)}}(E) & 0
\end{array} \right),
\end{equation}
on obtient $T_{\omega^{(0)}}(E)=\exp(\ell X_{\omega^{(0)}}(E))$.

\noindent Puis, notons $\lambda_1^{\omega^{(0)}},\ldots, \lambda_N^{\omega^{(0)}}$ les valeurs propres r\'eelles de la matrice r\'eelle sym\'etrique $M_{\omega^{(0)}}(0)$. Alors les valeurs propres de $X_{\omega^{(0)}}(E) ^tX_{\omega^{(0)}}(E) $ sont $1$, $(\lambda_1^{\omega^{(0)}}-E)^2$, $\ldots$, $(\lambda_N^{\omega^{(0)}}-E)^2$, donc $||X_{\omega^{(0)}}(E)||=\max(1,\max_{1\leq i\leq N} |\lambda_i^{\omega^{(0)}}-E|)$ o\`u $||\ ||$ d\'esigne la norme matricielle induite par la norme euclidienne  sur $\R^{2N}$.

\noindent Soit $\mathcal{O}$ le voisinage de l'identit\'e donn\'e par le th\'eor\`eme \ref{thm_BG} pour $G=\SpN$. Alors $\mathcal{O}$ ne d\'epend que de $N$. On pose :
$\dlO=\max \{ R>0\ |\ B(0,R) \subset \log\, \mathcal{O} \}$,
o\`u $B(0,R)$ d\'esigne la boule de centre $0$ et de rayon $R>0$ pour la topologie induite par la norme matricielle $||\ ||$ sur l'alg\`ebre de Lie $\spN$ de $\SpN$. On veut trouver un intervalle de valeurs de $E$ telles que :
\begin{equation}\label{ineg_1}
\forall \omega^{(0)}\in \{ 0,1\}^N,\ 0<\ell ||X_{\omega^{(0)}}(E)|| < \dlO,
\end{equation}
soit encore,
\begin{equation}\label{ineg_2}
0<\ell \max \left(1,\max_{\omega^{(0)}\in \{ 0,1\}^N} \max_{1\leq i\leq N} |\lambda_i^{\omega^{(0)}}-E|\right) < \dlO.
\end{equation}
Supposons que $\ell \leq \dlO$ et posons $r_{\ell}=\frac{1}{\ell}\dlO\geq 1$. On veut caract\'eriser l'ensemble
\begin{equation}\label{ens_I}
I_{\ell}=\left\{E\in \R\ \bigg|\  \max \left(1,\max_{\omega^{(0)}\in \{ 0,1\}^N} \max_{1\leq i\leq N} |\lambda_i^{\omega^{(0)}}-E|\right) \leq r_{\ell} \right\}.
\end{equation}
Comme $r_{\ell} \geq 1$, $I_{\ell}=\cap_{\omega^{(0)}\in \{ 0,1\}^N} \cap_{1\leq i\leq N} [\lambda_i^{\omega^{(0)}}-r_{\ell}, \lambda_i^{\omega^{(0)}}+r_{\ell}]$. Posons : 
{\small \begin{equation}\label{lmax_lmin}
\lambda_{\mathrm{min}}=\min_{\omega^{(0)}\in \{ 0,1\}^N} \min_{1\leq i\leq N} \lambda_i^{\omega^{(0)}},\ \lambda_{\mathrm{max}}=\max_{\omega^{(0)}\in \{ 0,1\}^N} \max_{1\leq i\leq N} \lambda_i^{\omega^{(0)}}\ \mathrm{et}\ \delta=\frac{\lambda_{\mathrm{max}}-\lambda_{\mathrm{min}}}{2}.
\end{equation} }
Alors, si $\delta < r_{\ell}$, $I_{\ell}=[\lambda_{\mathrm{max}}-r_{\ell},\lambda_{\mathrm{min}}+r_{\ell}]$ et $I_{\ell}$ est l'intervalle centr\'e en $\frac{1}{2}(\lambda_{\mathrm{min}}+\lambda_{\mathrm{max}})$ et de longueur $2r_{\ell}-2\delta>0$. De plus, $2r_{\ell}-2\delta$ tend vers l'infini lorsque $\ell$ tend vers $0$ et comme $\lambda_{\mathrm{min}}$, $\lambda_{\mathrm{max}}$ et $\dlO$ ne d\'ependent que de $N$, $I_{\ell}$ ne d\'epend que de $\ell$ et de $N$. La condition $\delta < r_{\ell}$ est \'equivalente \`a $\ell < \frac{\dlO}{\delta}=\ell_C (N)$.

\noindent Donc il existe $\ell_C= \frac{\dlO}{\delta}$ tel que pour tout $\ell < \ell_C$, il existe un intervalle compact $I(N,\ell)=[\lambda_{\mathrm{max}}-r_{\ell},\lambda_{\mathrm{min}}+r_{\ell}]$ (ne d\'ependant que de $\ell$ et de $N$ et dont la longueur tend vers l'infini lorsque $\ell$ tend vers $0$) tel que :
\begin{equation}\label{ineq4}
\forall \omega^{(0)}\in \{ 0,1\}^N,\ \forall E \in I(N,\ell),\ 0< \ell ||X_{\omega^{(0)}}(E) || \leq \dlO.
\end{equation}
Alors, pour tout $E\in I(N,\ell)$, $\log T_{\omega^{(0)}}(E) = \ell X_{\omega^{(0)}}(E) $ puisque $\exp$ est un diff\'eomorphisme de $\log \mathcal{O}$ sur $\mathcal{O}$. Or, on peut v\'erifier alg\'ebriquement que 
\begin{equation}\label{egal_lie}
\forall \ell >0,\ \forall E\in \R,\ \mathrm{Lie} \{ \ell X_{\omega^{(0)}}(E) \ |\ \omega^{(0)}\in \{ 0,1\}^N \}=\spN 
\end{equation}

\noindent (voir Proposition IV.5.12 dans \cite{CL90}). Alors, par le th\'eor\`eme \ref{thm_BG} on obtient :
\begin{equation}\label{groupe_eng}
\forall \ell < \ell_C,\ \forall E\in I(N,\ell),\ \overline{<T_{\omega^{(0)}}(E)\ |\ \omega^{(0)} \in \{ 0,1\}^N>}=\SpN .
\end{equation}
\vskip 1mm
\noindent Donc, comme $\overline{<T_{\omega^{(0)}}(E)\ |\ \omega^{(0)} \in \{ 0,1\}^N>}\subset G(E)$ et $G(E)\subset \SpN$, 
\begin{equation}\label{egal_final}
\forall \ell < \ell_C,\ \forall E\in I(N,\ell),\ G(E)=\SpN .
\end{equation}
Cela prouve la s\'eparabilit\'e des exposants de Lyapounov associ\'es \`a $H_{\ell}(\omega)$. L'absence de spectre absolument continu pour $H_{\ell}(\omega)$ dans $I(N,\ell)$ en d\'ecoule en utilisant la th\'eorie de Kotani et Simon (voir \cite{KS88,boumaza-rmp}).

%-------------------------------------------------------------%

\end{document}